\documentclass[iop]{emulateapj}

\usepackage{color}

\begin{document}
\title{The Fate of the Compact Remnant in Neutron Star Mergers}

\author{Chris L. Fryer\altaffilmark{1,2}, Krzysztoff
  Belczynski\altaffilmark{3}, Enrico Ramirez-Ruiz\altaffilmark{4},
  Stephan Rosswog\altaffilmark{5}, Gang Shen\altaffilmark{6}, Andrew
  W. Steiner\altaffilmark{7, 8}}

\altaffiltext{1}{Department of Physics, The University of Arizona,
  Tucson, AZ 85721} 
\altaffiltext{2}{CCS Division, Los Alamos National Laboratory, Los
  Alamos, NM 87545} 
\altaffiltext{3}{Astronomical Observatory, University of Warsaw, Al
  Ujazdowskie 4, 00-478 Warsaw, Poland}
\altaffiltext{4}{Department of Astronomy and Astrophysics, University
  of California, Santa Cruz, CA 95064}
\altaffiltext{5}{The Oskar klein Center, Department of Astronomy, AlbaNova, 
Stockholm University, SE-106 91 Stockholm, Sweden}
\altaffiltext{6}{Institute for Nuclear Theory, University of
  Washington, Seattle, Washington 98195}
\altaffiltext{7}{Department of Physics and Astronomy, University of
Tennessee, Knoxville, Tennessee 37996, USA}
\altaffiltext{8}{Physics Division, Oak Ridge National Laboratory, Oak
Ridge, Tennessee 37831, USA}

\begin{abstract}

Neutron star (binary neutron star and neutron star - black hole)
mergers are believed to produce short-duration gamma-ray bursts.  They
are also believed to be the dominant source of gravitational waves to
be detected by the advanced LIGO and the dominant source of the heavy
r-process elements in the universe.  Whether or not these mergers
produce short-duration GRBs depends sensitively on the fate of the
core of the remnant (whether, and how quickly, it forms a black hole).
In this paper, we combine the results of merger calculations and
equation of state studies to determine the fate of the cores of
neutron star mergers.  Using population studies, we can determine the
distribution of these fates to compare to observations.  We find that
black hole cores form quickly only for equations of state that predict
maximum non-rotating neutron star masses below 2.3-2.4 solar masses.
If quick black hole formation is essential in producing gamma-ray
bursts, LIGO observed rates compared to GRB rates could be used to
constrain the equation of state for dense nuclear matter.

\end{abstract}


\section{Introduction}
\label{sec:intro}

Since their discovery in the late 1970s~\citep{klebesadel76},
scientists have proposed a growing number of progenitors and engines
for gamma-ray bursts (GRBs).  The variety in the observations of
classical GRBs points to multiple progenitor scenarios.  Based on
their duration, bursts are traditionally separated in long and short
GRBs~\citep{kouveliotou93}.  Within the black hole accretion disk
(BHAD) class of models, long-duration bursts are believed to arise
from accreting black holes at the center of massive stars, e.g.
from collapsars~\citep{woosley93} or he-mergers~\citep{fryer99}.
Short-duration bursts are believed to be produced by the merger of two
compact objects: consisting either of a binary neutron star (NS-NS) or
a neutron star and a black hole (BH-NS): for review see, for example,  
\cite{fryer99,popham99,lee07,rosswog15a}.  These compact mergers are
produced in close binaries and the kicks imparted onto these compact
remnants cause these systems to have high space velocities.  These
velocities, coupled with merger times, mean that bursts produced from
NS-NS and BH-NS binaries should have broad spatial distributions with
respect to their host galaxy, including bursts that occur well outside
of their host~\citep{bloom99,fryer99,zemp09}.  With Swift, the number
of localized short-duration bursts has increased dramatically, and the
observed spatial distributions of these bursts matches the predictions
from theory~\citep{belczynski06,fong13,behroozi14}.  Indeed, no other
model to date can easily explain the spatial distributions of
short-duration bursts, and NS-NS and BH-NS mergers are almost
universally considered the leading progenitors for these bursts.

The BHAD engine is not the only way accreting compact objects can
produce possible outflows.  Neutron stars are also able to accrete at
Super-Eddington rates~\citep{houck91,fryer96} and NS accretion disk (NSAD)
systems will look very similar to BHAD engines.  Alternatively,
building off of the leading mechanism for soft gamma-ray repeaters
(SGR), theorists have argued that rapidly-spinning
magnetars~\citep{duncan92} can produce classical GRBs~\citep{zhang01}.
The disadvantage of the neutron star models is that they -- as long as
the merger remnant is stable -- can drive strong baryonic winds via
neutrino energy deposition \citep{dessart09,perego14} that
potentially choke the GRB jet~\citep{rosswog02,murguiaberthier14}.

The advantage of the neutron star model is that the NS can be born
with strong magnetic fields\footnote{Note that generally in high
  accretion scenarios like these merger systems, there is a belief
  that the accretion buries the field\citep{popov12,vigano12}.
  Typical timescales for the re-emergence of these magnetic fields are
  tens of kyr.  In such a case, any magnetar-like model will not work
  to explain a GRB.}  and these magnetars provide a means to drive
late-time emission~\citep{usov92,thompson94,rowlinson14}.  Understanding the relative
rate of merging systems that form black holes versus neutron stars can
help direct theorists toward a better understanding of these systems.
For example, if very few systems collapse to form black holes, either
the baryonic loading problem is less severe than currently believed or
mergers are not the solution to GRBs.  On the other extreme, if only a
small fraction remain neutron stars, it is worth identifying and
trying to observe the failed GRBs formed by these systems.
Observations of relative rates of these systems can provide a fairly
direct means to constrain the nuclear equation of state.

This paper brings together studies of binary neutron star merger
models, nuclear equations of state, and population synthesis models to
study the formation rate of different potential GRB progenitors,
differentiating between black hole accretion disk models,
accretion-induced collapse models, and neutron star accretion models.
Section~\ref{sec:mergermodels} describes the merger models used in
this study and Section~\ref{sec:eos} describes the effects and
uncertainties in the equation of state (EOS).  By coupling these
results with population synthesis studies, we can compare these
results to the suite of GRB observations (Section~\ref{sec:popsynth}).
We conclude with a discussion of the implications of these results on
the wide range of phenomena explained by NS-NS and BH-NS binaries.

\section{Merger Models}
\label{sec:mergermodels}

In the past decade, the number of hydrodynamic simulations of NS-NS
and BH-NS mergers has grown
considerably\citep{foucart10,foucart11,foucart12,korobkin12,hotokezaka13,kyutoku13,takami14,bauswein14b,radice14,shibata14,kiuchi14,foucart15}.
Although these models are becoming increasingly sophisticated, most of
the current models include only a subset of the physics needed to
model these objects: hydrodynamics, neutrino transport (or at least
neutrino energy losses), nuclear equations of state, magnetic fields,
and general relativity.  Nevertheless, these models are gradually
painting a complete picture of the merger process and we can use them
to make a first pass at the fate of these systems.

\subsection{BH-NS mergers}
\label{sec:bhns}

To date, no BH-NS system has been observed and, at this point, there are no
observations that would require such systems to exist.  However, there are
clear observational biases against such systems: if the NS is formed
after the black hole, as we expect, it is difficult to recycle it.
Hence we do not expect a millisecond pulsar in BH-NS binaries.
Population synthesis models predict that these systems form frequently
and many of them close enough to merge within a Hubble time. 
Theoretical models for massive star evolution
and the formation of BH-NS systems suffer from large uncertainties
(e.g., treatment of star internal mixing, stellar winds, common
envelope and supernova explosion).  Population synthesis models
predict a range of formation rates from several orders of magnitude
below the BNS merger rate to rates that rival the BNS
rate~\citep{dominik12}.

The binary parameters of BH-NS systems are crucial for the outcome 
of the merger and only for a fraction of the parameter space will it be
possible to form an accretion disk massive enough to launch a GRB.
If we assume the energy arises from the disk, 
\begin{equation}
E_{\rm ext} \sim 2\times 10^{51} {\rm erg} \left( \frac{\epsilon}{0.1}\right) 
                                                                \left( \frac{M_{\rm disk}}{0.01 M_\odot} \right)
\end{equation}
can be extrated, so that the disks do not need to be extremely massive
to accomodate the typical isotropic gamma-ray energies of short
bursts~\citep{berger14} of $\sim 10^{50}$ erg (and correspondingly
lower if they are collimated).  If the energy arises from the rotating
black hole, significantly smaller disk masses are needed to produce
the energies required for short bursts~\citep{lee07}.  Nevertheless,
not all systems will be able to form disks, since the radius where
tidal disruption sets in, $R_{\rm tid}$, must be larger than the
innermost stable circular orbit (ISCO), $R_{\rm ISCO}$.  Since $R_{\rm
  ISCO} \propto M_{\rm BH}$, but $R_{\rm tid} \propto M_{\rm
  BH}^{1/3}$, for massive enough black holes the tidal radius lies
inside of the ISCO and forming a massive accretion disk becomes
impossible. For a non-spinning black hole this occurs already near
$M_{\rm BH}= 8$ M$_\odot$, so that BHs of the masses that are
thought~\citep{belczynski08a,ozel10} to be most likely ($\sim 10$
M$_\odot$) need large dimensionless spin parameters, $a_{\rm BH}
\gtrsim 0.9$, to form sizeable disks \citep{foucart12}.  Finally, the
inclination of vector of BH-NS orbital angular momentum to BH spin
vector plays an important role in the formation of a torus.  At large
inclination angles ($\gtrsim 40-90^\circ$), material from the
disrupted NS is ejected from the vicinity of a merger and does not
form torus. Non-zero tilts are expected in BH-NS systems if a NS is
formed as a second compact object in a binary and with a non-zero
natal kick in isolated binary evolution in field
populations~\citep{fryer99,dominik12,dominik13,dominik14} and, in
majority of cases, in dynamical evolution; e.g., globular
clusters\citep{grindlay06,lee10,east13,rosswog13,tsang13,ramirezruiz15}.

The number of simulations studying these mergers and their fates has
grown with time \citep{rosswog04,rosswog05,rantsiou08,foucart10,foucart11,kyutoku13,paschalidis14,foucart15}, all showing that the final fate of the merged system
depends sensitively upon the initial conditions. Based on earlier merging BH-NS
models by \cite{rantsiou08}, it was shown for one particular
evolutionary model that only a fraction of BH-NS systems may
potentially form an accretion torus and lead to a GRB. For small BH
spins ($a_{\rm spin}<0.6$), only 1\% of BH-NS mergers were found to
have a disk needed to produce a GRB.  For high BH spins the fraction
of systems with disks have been calculated to be $\sim
40\%$~\cite{belczynski08a}.  It is important, that in future studies,
the merger hydrodynamical simulations are mapped with
astrophysically-motivated predictions of BH-NS system physical
properties.

For this paper, we will assume the GRB rate is dominated by BNS
mergers, but we will return to the topic of BH-NS mergers when we 
better understand the BNS results.

\subsection{NS-NS mergers}
\label{sec:nsns}

NS-NS mergers always form sizable accretion disks and aare the more
canonical GRB model.  The primary difficulty with these mergers as GRB
progenitors is the nature of the merged core.  After the merger, the
remnant will evolve according to four separate pathways: collapse
directly to black holes, those that initially form NSs but
subsequently collapse during disk accretion, those that don't collapse
to black holes until after the disk has fully accreted and the newly
formed neutron star spins down, and those that, even after the spin
down, remain a NS.  To determine the fate of each merged object, we
must combine both simulations of the merger process (discussed in this
section) with our current understanding of the nuclear equation of
state (see Section~\ref{sec:eos}).  For two equally-massed neutron
stars, the merger produces a condensed core consisting of two major
components formed from the cores of the initial neutron stars.  As the
mass ratio becomes more extreme, the more massive neutron star remains
more intact, disrupting its companion.  In all cases, the merger
produces a merged core surrounded by a dense accretion disk with a
small amount of material ($\sim$1\%) ejected during the tidal ejection
process.

To study the fates of neutron stars, we use the large grid of smooth
particle hydrodynamics simulations from \cite{korobkin12}.  This grid of models
includes a broad suite of neutron star mass pairs using the same input
physics.  For each simulation, we calculate the mass and angular
momentum of the merged core (all material above $10^{14}\,{\rm g
  cm^{-3}}$), disk masses (bound mass with densities below
$10^{14}\,{\rm g cm^{-3}}$) and ejecta masses
(Table~\ref{tab:merger}).

At the end of the smooth particle hydrodynamics simulations,
considerable mass ($\sim 0.5-0.9\,M_\odot$) remains bound but at
densities below $10^{14}\,{\rm g cm^{-3}}$.  In
Table~\ref{tab:merger}, we refer to this material as ``disk'' material
and much of it will form a disk that quickly accretes onto the compact
core.  If the core collapses to a black hole, the system immediately
evolves into a BHAD engine.  If not, it may first
pass through a neutron star accretion disk or magnetar engine until
the accretion drives the neutron star above the maximum mass and it
collapses, producing a BHAD scenario.

Whether or not the core is a black hole or neutron star depends upon
whether it is more massive than the maximum neutron star mass at
its spin rate.  This depends upon the still-uncertain equation of
state.  Recent observations by \cite{demorest10} and
\cite{antoniadis13} demonstrate that this mass is least 2\,M$_\odot$.
The exact value for the maximum mass of a particular merged core
depends on the equation of state as well as the internal energy and
angular momentum distribution in the core.  Our choice of equation of state will
determine which merged cores collapse immediately to a black hole.

\begin{figure}[!hbtp]
\centering
\includegraphics[width=\columnwidth]{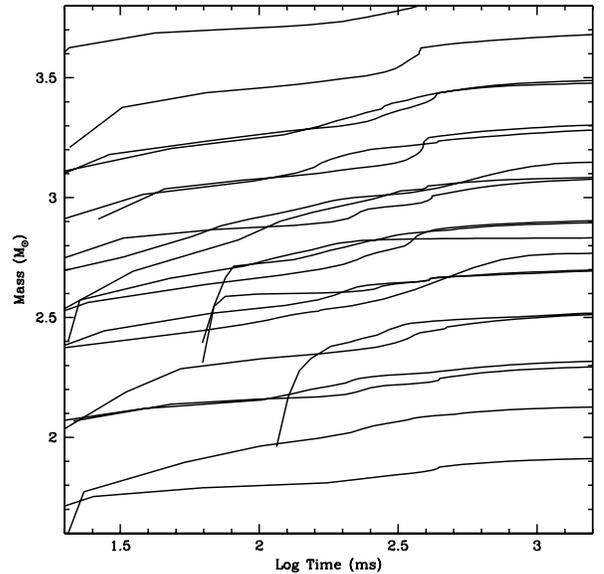}
\caption{Core mass versus time for our merged systems assuming 
an $\alpha$-disk solution for the accretion time.  Note that 
in many systems, the accretion occurs very rapidly (in the first 
10,ms) and can last out to a few seconds.  The jumps occur because 
the angular momentum of material is not uniform and large amounts 
of material have similar angular momenta, and hence, accretion 
timescales.}
\label{fig:acc}
\end{figure}

In other cases, the core must accrete some material before it
collapses.  To determine the engine behind this neutron star, we must
understand the accretion timescale.  If the core remains a neutron
star for a long period of time, the neutrino-driven wind will choke
the outflow, preventing a strong burst~\citep{murguiaberthier14}.  To
estimate the accretion time of the disk, we use the matter
distribution from our models and assume an $\alpha$-driven disk
scenario.  That is, for the spatial distribution of the disk material,
we assume the accretion timescale is just the orbital period divided
by the value of $\alpha$: $T_{\rm acc}=\frac{1}{\alpha}\frac{2 \pi
  r^{3/2}}{(G M_{\rm encl})^{1/2}}$ where $r$ is the spatial position
of the matter, $M_{\rm encl}$ is the enclosed mass and $\alpha$ is the
viscous disk parameter.  Figure~\ref{fig:acc} shows the mass of the
core as a function of time under our $\alpha$-disk assumptions using
$\alpha=0.01$.  Note that although much of the accretion occurs in the
first 100\,ms, the accretion phase can last out to a few seconds for
some systems.  Clearly, the accretion time depends linearly on the
value for $\alpha$ whose value is uncertain to an order of magnitude.

For the most part, these models assume that the neutron stars in the
binary are initially not spinning (although for one case, we include a
neutron star binaries in co-rotation).  The assumption is well-justified
since the last inspiral stages happen so quickly that even an
unplausibly large dissipation cannot spin up the stars
substantially~\citep{bildsten92,kochanek92}.  Pre-merger 
spins are typically substantially smaller than co-rotation, but even in the co-rotation 
case, the mass is  altered by only a few percent, but the angular momentum in the
core can change by 15\%.  This has only a small affect on the maximum 
stable neutron star mass of the core, but it can dramatically change
the accretion rate (see Section~\ref{sec:eos}).

Be aware that the current set of simulations are incomplete, missing
both physics and spatial resolution to model the physics.  As this
physics is improved, exact quantities on merged cores, etc. will move.
But the basic trends will continue and, although there may be a shift
in our results, the trends and methodology to study NS-NS and BH-NS
mergers the results will remain.

\section{Equation of State}
\label{sec:eos}

The fate of the core depends on the exact details of the equation of
state.  If the maximum neutron star mass were 2.0M$_\odot$, the
maximum observed neutron star mass~\citep{demorest10,antoniadis13}, most of our
merger cases would immediately collapse to a black hole.  This
observation only places a lower limit on the maximum mass, and the
maximum mass of a cold neutron star may, depending upon the equation
of state, be much higher than this maximum observed mass.  In
addition, fast-rotation and thermal energy can significantly raise the
maximum mass.  To determine the final fate of our merger remnants, we
will study the maximum mass, and its dependence on the spin and
internal energy, for a range of equation of state models.

The merged core, especially if the components are near-equal mass, are
often born with strong differential rotation~\citep{rosswogdavies02}.
Depending upon the magnetic field strength, these cores will very
quickly evolve into uniform rotation.  For example, \cite{shibata07}
argue that a combination of magnetic winding, shearing and the
magnetorotational instability can quickly redistribute the angular
momentum in an accreting disk system.  For a few of our fastest
rotating mergers (the merger of a 2.0\,M$_\odot$ NS with a
1.0,1.2\,M$_\odot$ NS and the merger of a 1.8 \,M$_\odot$ NS with a
1.0\,M$_\odot$ NS: see Table~\ref{tab:merger}), secular instabilities
can develop that will also redistribute the angular momentum.
Finally, neutrino cooling in the core will drive convection in the
core~\citep{rosswog03} that can redistribute the angular momentum.
These redistribution mechanisms have timescales lying between
10-100\,ms.  If it is at the low end, our assumption that it is nearly
instantaneous holds.  If there is a longer delay in this evolution to
uniform rotation, the maximum mass could remain high for longer,
producing more long-lived neutron star systems.  Note, however, that
for our population study (see section~\ref{sec:popsynth}), only
$0.3$\% of our systems have a NS component with masses above 1.8
\,M$_\odot$ and none of our $\sim 6000$ systems included pairs with a massive
component above 1.8 \,M$_\odot$ and a low mass component below 1.2
\,M$_\odot$, so the highest angular momentum systems are extremely rare.

Although a multitude of dense matter equations of state (EOS) exist
which make firm predictions about maximum masses, see for example,
\cite{cook94}, all of them make simplifying assumptions for the
physics and theory can not place strong constraints on the final
maximum mass.  Most models for the neutron star equation of state
predict masses in a range between 2.0 and 3.0M$_\odot$,
uncertainties in nuclear physics prevent a more precise prediction of
the maximum mass for a non-rotating model.  However, for a model with
a given maximum non-rotating mass, we can predict the effect of
thermal energy and angular momentum.

For our study, we use a variety of equations of state to probe the
fate of our merged cores.  These include, inparticular, the
NL3\citep{lalazissis97} and FSU2.1~\citep{toddrutel05} EOSs. We also
include one of the equation of state models developed
in~\citet{steiner10,steiner13,steiner15} where the model parameters
are matched to observations of quiescent low-mass X-ray binaries and
photospheric radius expansion bursts~\citep{steiner15} using a Monte
Carlo scheme to perform marginal
estimation~\citep{steiner14,steiner14b}. In this equation of state
model, matter near and below the nuclear saturation density is
described by a parameterization calibrated to the properties of
laboratory nuclei and quantum Monte Carlo
simulations~\citep{steiner12}. At higher densities, matter is
described by two piecewise polytropes. The RNS code for rotating
neutron stars from \citet{stergioulas95} was embedded in our Monte
Carlo simulation to output probability distributions for the maximum mass of a
neutron star given a fixed angular momentum.

During the merger, shocks and viscous forces increase the energy in the
core, raising the entropy.  This extra thermal energy can raise the
maximum neutron star mass~\citep{kaplan14,bauswein14a}. Most studies
where this has been a large effect have assumed entropies of roughly
8\,$k_{\rm B}$ per nucleon.  However, actual estimates of the
post-merger entropies used in our study~\citep{korobkin12} lie at
about 1\,$k_{\rm B}$ per nucleon (Table~\ref{tab:merger}).  Using the 
NL3\citep{lalazissis97} equation of state (fairly typical) an entropy
of 1.28\,$k_{\rm B}$ per nucleon raises the maximum stable mass by
only 1\%.  In this scenario, the internal energy deposited in the core
in the merged system is not sufficient to strongly affect the maximum
neutron star mass.  There is more thermal energy in the surrounding
disk, but its low densities allow efficient neutrino cooling that is
on par with the accretion time.  Hence, we will use the accretion time
to estimate our collapse rate for systems that collapse after
accretion.

\begin{figure}[!hbtp]
\centering
\includegraphics[width=\columnwidth]{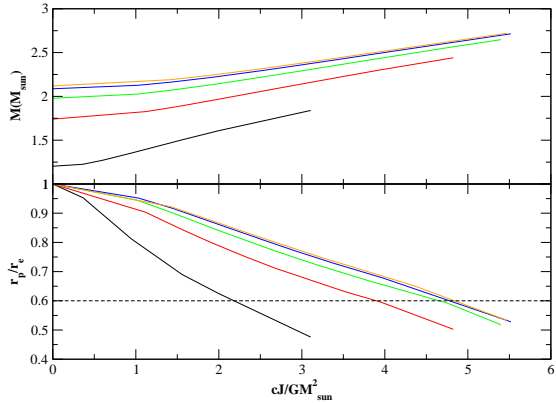}
\caption{Maximum neutron star mass versus angular momentum for the 
  NL3 equation of state for a set of central pressures (top).  A neutron
  star with a maximum non-rotating mass of 2.1\,M$_\odot$ will have a
  maximum mass of 2.6\,M$_\odot$.  This angular momentum can
  drastically distort the neutron star.  The ratio of the polar to
  equatorial radii (assuming solid-body rotation) is also shown as a
  function of angular momentum.  At the end of our simulations, the
  core structure has not yet reached solid body rotation, and the
  asphericity is even more extreme (possibly arguing for a more
  extreme maximum mass).}
\label{fig:shen}
\end{figure}

The angular momentum in the merged core can, however, have a larger
impact on the maximum neutron star mass.  If we assume the core
quickly evolves into a system in solid body rotation, we can use
existing equations of state to estimate the maximum neutron star mass
for these rotating systems.  For the FSU2.1 equation of state, we find
that the maximum neutron star mass increases nearly linearly with
rotational support over the range of rotations produced in neutron
star mergers (Fig.~\ref{fig:shen}).  As input for our population
studies, we have fit the maximum neutron star mass predicted for this
equation of state with the following linear approximation:
\begin{equation}
M_{\rm max,NS} = (2.1+J/(10^{50} {\rm g \, cm^2 \, s^{-1}})) M_\odot
\label{eq:nsm1}
\end{equation}
where $J$ is the rotational angular momentum of the compact core.
Figure~\ref{fig:shen} also shows the distortion of the neutron star 
that would form with such high angular momenta and this distortion 
reflects the large asymmetries produced in the cores of these merger 
events.

\begin{figure}[!hbtp]
\centering
\includegraphics[width=\columnwidth]{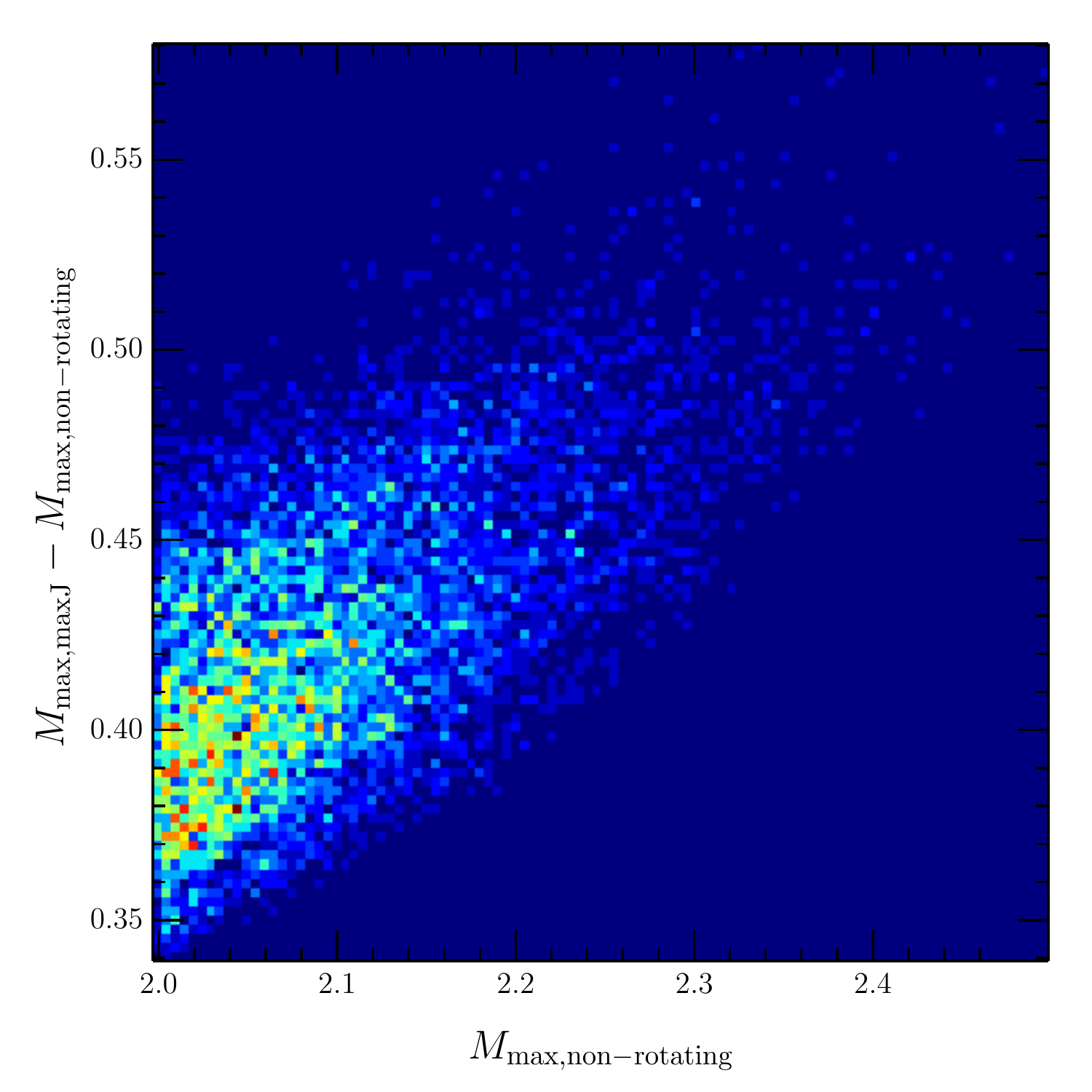}
\includegraphics[width=\columnwidth]{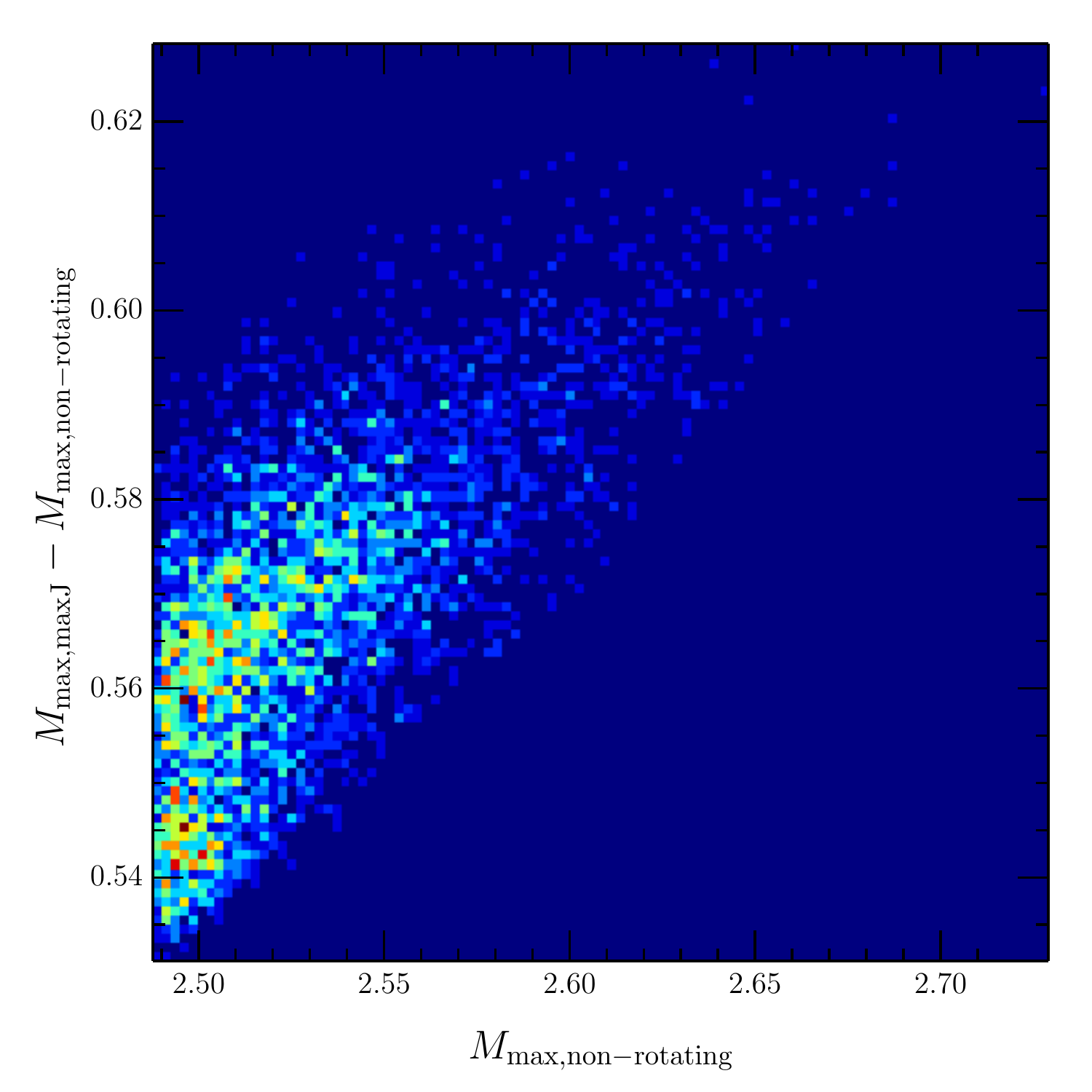}
\caption{Probability distribution of $\Delta M (M_\odot)$ as a
  function of the maximum non-rotating mass.  The top panel focuses
  on masses near and above a maximum mass of 2.0\,M$_\odot$ whereas
  the bottom panel zooms in on those maximum non-rotating masses above
  2.5\,M$_\odot$.}
\label{fig:steiner}
\end{figure}

Thus far we have only used specific equations of state.  To determine
the role of angular momentum change with a broad range of allowed
nuclear equations of state, we use our parameterized equation of
state.  The increase in the mass ($\Delta M$) of a maximally-rotating
neutron star depends upon the maximum mass of the non-rotating neutron
star.  Figure~\ref{fig:steiner} shows the distribution of $\Delta M$
as a function of maximum non-rotating mass.  For a $2.0\,M_\odot$
star, a maximally-rotating neutron star lies between
$2.35-2.5\,M_\odot$ depending upon the exact equation of state.  For
an equation of state producing a maximum non-rotating mass of
$2.6\,M_\odot$, this mass is $3.16-3.22\,M_\odot$, an increase of
$0.56-0.62\,M_\odot$.  The $\Delta M$ for a maximally rotating neutron
star increases for equations of state that have larger maximum
non-rotating stars.  However, for the angular momenta in our merged
cores, the increase in the maximum mass is much less dramatic.  A fit
to these lower-angular momentum systems yields the following result
for the maximum rotating neutron star mass:
\begin{equation}
M_{\rm max,NS} = M_0 + M_1\times (J/(10^{49} {\rm g \, cm^2 \, s^{-1}}))^{\beta} M_\odot
\label{eq:nsm2}
\end{equation}
where $M_0$ is the maximum mass for a non-rotating neutron star.  For
this paper, we study different values of $M_0$, fitting the change in
mass to average values of rotating systems for a range of angular
momenta from $0-3\times10^{49}{\rm g \, cm^2 \, s^{-1}}$ based on the
angular momenta in the cores in our simulations
(Table~\ref{tab:merger}).  For $M_0=2.0,2.5,2.7\,M_\odot$, the values
for $M_1$ and $\beta$ are: $M_1=0.0219,0.0354,0.0574$ and
$\beta=1.75,1.5,1.6$ respectively.  Typically, rotation adds less than
a few tenths of a solar mass to our maximun mass of our merged cores.
Accretion of the disk material can also alter the angular momentum of
the core, but how much angular momentum depends upon the exact
accretion process (including magnetic field strength).  For this
study, we assume the specific angular momentum is held constant during
the accretion phase.  Moreover, we assume that the post-merger
magnetic fields are small enough ($B \|| 10^{17}$G) to not contribute
noticeably to the pressure.  To obtain a more detailed survey of the
possible merger fates, we created a grid of maximum masses, varying
$M_0$ from 2.0 to 2.5 with constant values for $M_1$
($=0.03$M$_\odot$) and $\beta$ ($=1.6$).

With these fits to the maximum mass of our equation of state, we can
determine the fate of the merged systems presented in
Section~\ref{sec:mergermodels}.  Here we use the four fates discussed
in Section~\ref{sec:nsns}: BH, BH$_{\rm acc}$, BH$_{\rm spin}$, NS.
Note that for these fates, our equation of state models assume that
the neutron star has reached an equilibrium density state and it is in
solid body rotation.  The rotating object is differentially rotating
through this period but our simplified assumptions provide an estimate
of the importance of spin in the final fate.  Spin may prevent an
immediate collapse to a black hole but, in most cases, the accretion
of the surrounding material is sufficient to cause the core to
collapse with or without rotation (see Table~\ref{tab:nsfate}).

\section{Tying it all together:  Population synthesis}
\label{sec:popsynth}

To compare these results to observations, we must incorporate our
findings into a study of the binary neutron star and neutron
star/black hole populations.  We employ the {\tt StarTrack} population synthesis code 
\citep{belczynski02,belczynski08b} to generate a population 
of binary compact objects (in particular NS-NS and BH-NS binaries). 
The code is based on revised formulas from \cite{hurley00}; updated with 
new wind mass loss prescriptions, calibrated tidal interactions, physical 
estimation of donor's binding energy ($\lambda$) in common envelope
calculations and convection driven, neutrino enhanced supernova engines. 
A full description of these updates is given in \cite{dominik12}. The two 
most recent updates take into account measurements of initial parameter 
distributions for massive O stars \citep{sana12} as well as a correction 
of a technical bug that has limited the formation of BH-BH binaries for 
high metallicity (e.g., $Z=0.02$).

Two major factors shape the final compact object mass: wind mass loss and 
core collapse/supernova compact object formation. For wind mass loss we use 
O/B type winds from \cite{vink01} and for other evolutionary stages (e.g., 
LBV winds) formulae as calibrated in \cite{belczynski10}. We adopt set of 
models presented by \cite{fryer12} with the rapid core collapse/supernova 
mechanism. The explosion occurs within the first $0.1-0.2$s driven by a 
convection and neutrino enhanced engine. This ``rapid'' supernova engine 
reproduces \citep{belczynski12} the observed mass gap (apparent lack of 
neutron stars and black holes with mass in range $2$--$5$\,M$_\odot$) in Galactic 
X-ray binaries~\citep{ozel10, bailyn98}. Additionally, we include NS star formation 
through electron capture supernovae (ECS) for low mass progenitors 
$M_{\rm zams} \sim 7$--$11$\,M$_\odot$. The range of initial progenitor mass on Zero 
Age Main Sequence (ZAMS) for ECS depends sensitively on interactions (mass loss 
and gain) of stars in binary. We assume that all NSs formed through ECS have
mass of $1.26$\,M$_\odot$. Stars initially more massive than $\sim 10$\,M$_\odot$ typically
form NS in Fe core collapse through our adopted rapid supernova engine. NS
are formed in a broad mass range $1.1$--$2.5$\,M$_\odot$, although most NSs that
form close NS-NS binaries are found in much narrower range $1.2$--$1.4$\,M$_\odot$
with a long tail extending to $2$\,M$_\odot$. In our evolutionary calculations we
assume that ECS NSs do not receive natal kicks \citep{podsiadlowski04}, while 
for NS formed in Fe core collapse we adopt a 1-D Maxwellian with 
$\sigma=265$ km s$^{-1}$ (that corresponds to an average kick of 
$\sim 400$ km s$^{-1}$ \citep{hobbs05}. 

To generate the populations of double compact objects we adopt power-law 
initial mass function (IMF) with slope of $-2.7$ for the massive primaries 
($M_{\rm zams}=5$--$150$\,M$_\odot$). Secondary mass (less massive binary component) 
is drawn from flat mass ratio distribution. Then we adopt two very different
models for initial orbital separation and eccentricity. In one set of models
we adopt flat in logarithm ($\propto 1/a$) distribution of initial separations 
with thermal eccentricity distribution ($=2e$). Such assumptions have been
adopted in many population synthesis studies (e.g., for references and
examples see \cite{belczynski08b}). We refer to the models employing these
assumptions as ``OLD''. It was recently pointed out that these distributions may
not be adequate for massive progenitors of NSs and BHs. We therefore adopt
revised distributions: closer orbital periods $\propto \log(P)^{-0.55}$ and
less eccentric orbits $\propto e^{-0.42}$~\citep{sana12}. We refer to
models with these initial conditions as ``NEW''. 

Each initial condition calculation is performed for two characteristic
stellar metallicities: solar composition $Z=Z_\odot=0.02$ and $Z=0.1
Z_\odot=0.002$. We refer to these models as high and low-metallicity models.  
In our mapping of population synthesis calculations with NS-NS merger
calculations, we use evenly mixed ($50\%$ --$50\%$) population of low and
high metallicity models. This is to mimic (in very approximate way) a
stellar content of Universe at various redshifts. 

In each calculation we allow for a different outcome of Roche lobe overflow
(RLOF) initiated by a Hertzsprung gap (HG) donor. If RLOF appears to proceed on a
dynamical timescale we perform CE envelope energy calculation to check
whether a given binary survives the event (model A). Under this prescription, 
we allow for the efficient formation of close double compact object binaries. 
For contrast, we assume that each event of dynamical RLOF with HG donor does
not lead to the formation of close double compact object (model B). Some of
HG donors (right after main sequence) may not have developed clear
core-envelope structure and are a subject to CE merger (e.g.,
\cite{belczynski07}) or at later times HG donors may not have yet
developed convective envelope and instead of CE they may potentially
initiate thermal-timescale mass transfer that does not allow for significant
orbital contraction and close double compact object formation.

Additionally, we allow for uncertainties in our calculation of NS mass. In
particular we allow NS mass to increase by a fixed factor and in particular
we chose this factor to be $0.1$\,M$_\odot$.   Predictions of neutron star masses 
fold in a number of uncertainties from understanding the mass of the core 
at bounce, the duration of the engine phase and the amount of fallback.  
At the low mass end, fallback is not a concern and the duration of the engine 
must be short~\citep{fryer12}.  But this initial NS mass can be larger by as much 
as 0.1\,M$_\odot$.  In many cases, the first neutron star formed also goes 
through a common envelope phase where it can accrete up to $\sim$0.1\,M$_\odot$ through 
hypercritical accretion~\citep{fryer96,belczynski02,belczynski10b,macleod15,macleod15b}.

\begin{figure}[!hbtp]
\centering
\includegraphics[width=\columnwidth]{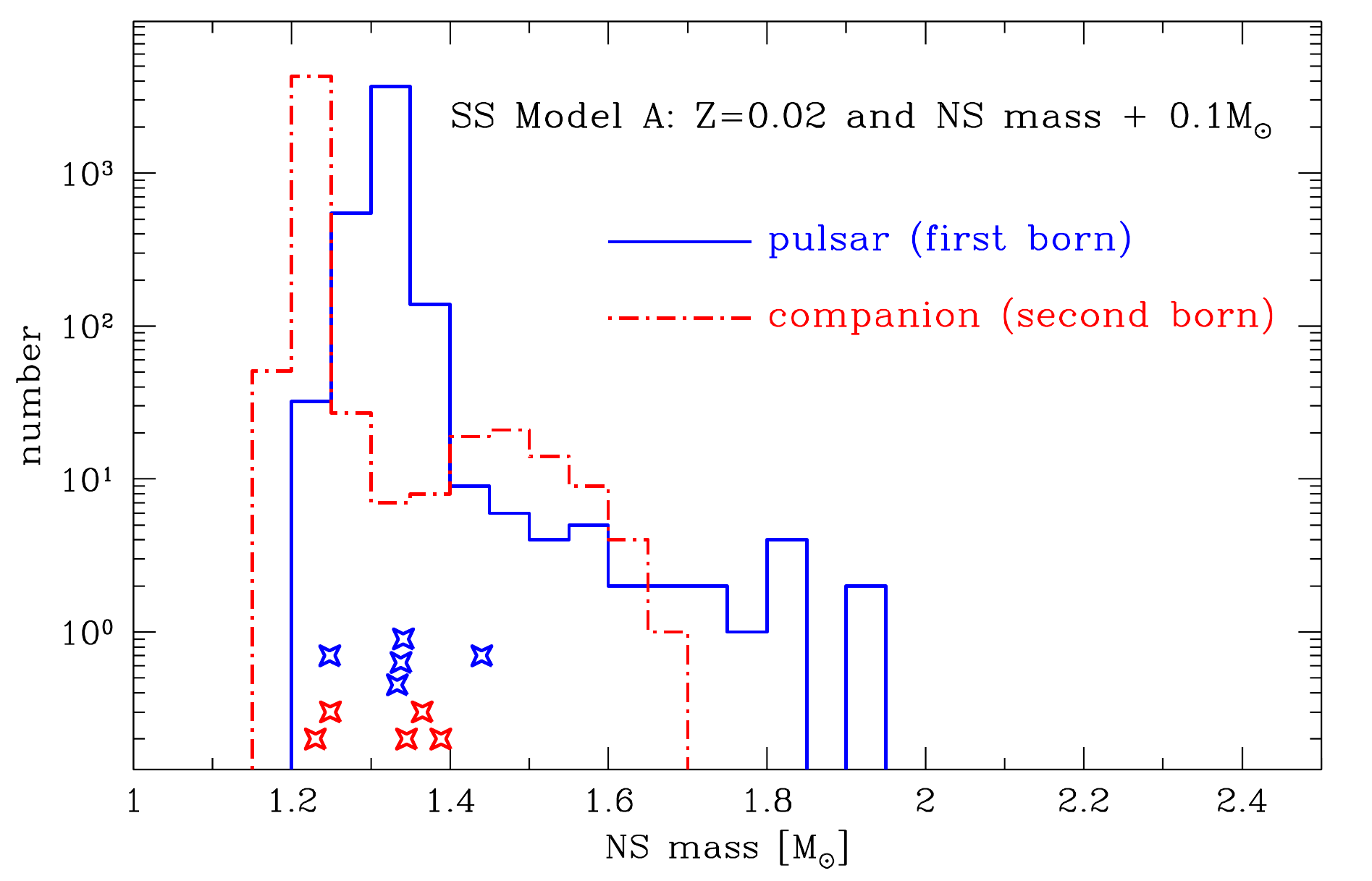}
\includegraphics[width=\columnwidth]{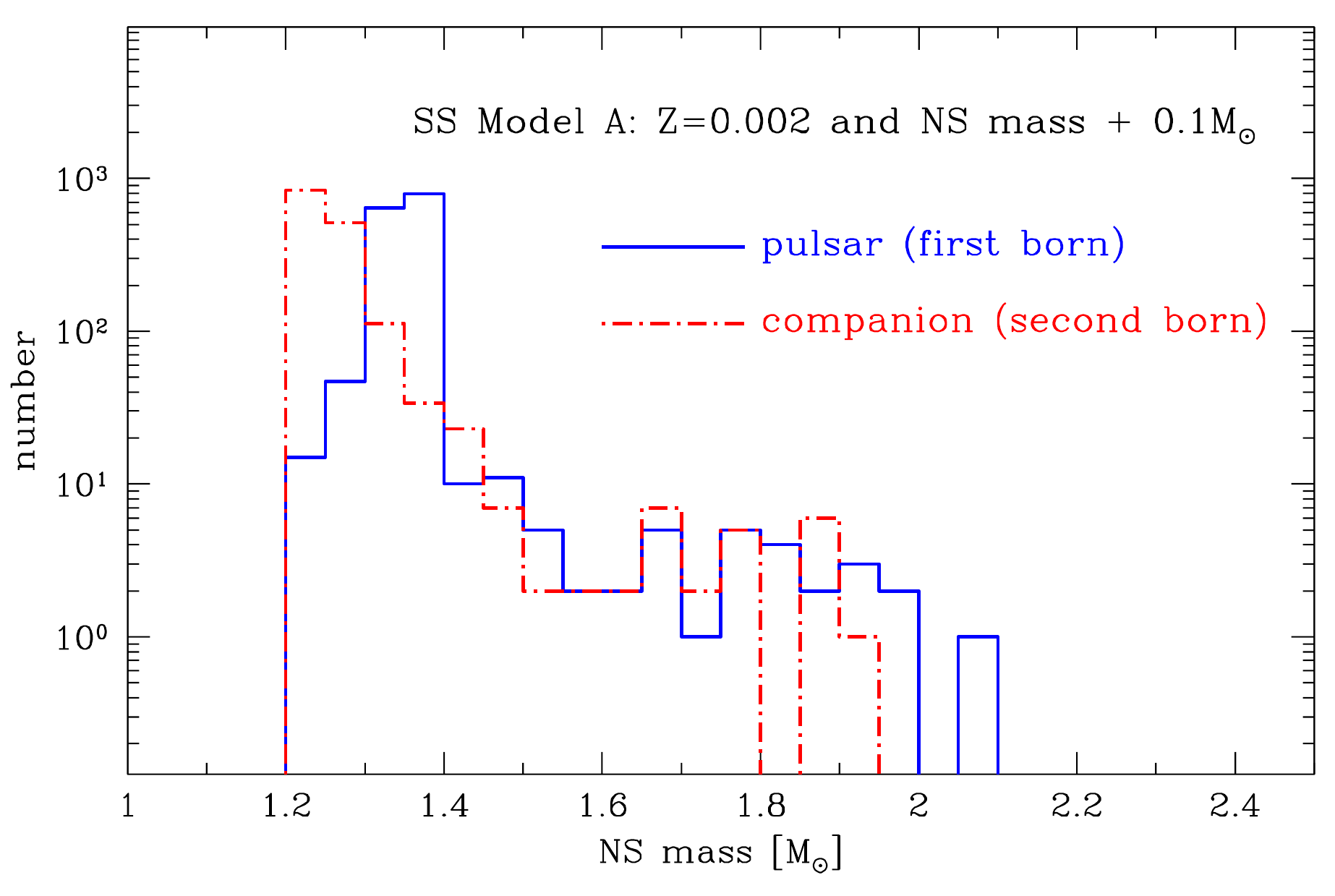}
\caption{The neutron star mass distribution (number per mass bin as a
  function of mass) for our models using the ``New'' initial
  conditions.  These results include systems that have undergone
  common envelope evolution in the Hertzprung gap with an enhanced
  neutron star mass (increasing it by 0.1\,M$_\odot$).  For
  comparison, we plot the masses of the observed NS-NS systems.}
\label{fig:nsmass}
\end{figure}

Physical properties of double compact objects of all sorts (NS-NS, BH-NS and
BH-BH systems) generated under above conditions are discussed in detail in a
series of recent papers \cite{dominik12}, \cite{dominik13}, \cite{dominik14} 
and de Mink \& Belczynski (2015, to be submitted). The actual population
synthesis models are available at {\tt www.syntheticuniverse.org}. 
In particular, generated NS-NS merger rates ($\sim 10$--$40$ Myr$^{-1}$) are in 
agreement with the empirically estimated Galactic NS-NS merger rates 
($\sim 10$--$100$ Myr$^{-1}$; \cite{kim15}). The NS masses predicted by our
models are in general agreement with the mass estimates available from close
NS-NS systems (Fig.~\ref{fig:nsmass}).  However, note that the X-ray binary 
4U1538-52 \cite{vankerkwijk95} suggests that neutron stars are born with mases 
below 1.1\,M$_\odot$, suggesting the minimum neutron star mass can not be 
much higher than our standard value, echoing the theoretical error bars of 
0.1M$_\odot$.

With these models, we can determine the fate distributions of merging
systems (if and when they form black holes) as a function of the
equation of state.  Table~\ref{tab:popsynthfate} shows the results for
a a series of models: with and without systems that undergo a common
envelope in the Hertzprung gap, with the standard initial conditions
and the new~\citep{sana12} initial conditions, with our standard and
modified (increased by 0.1\,M$_\odot$) neutron star masses.  The
general trends between all these models are the same.  The primary
fates of the merged binaries are bimodal: either the system collapses
to form a black hole after some disk accretion or the core remains a
neutron star.  Most of the systems that collapse to black holes do so
within the first 100\,ms.  This means that most of the black-hole
forming systems could also produce GRBs (avoiding the baryon
contamination problem).

The fraction of systems that collapse to a black hole is above 90\%
for any EOS that has a maximum non-rotating mass below
$\sim$2.3-2.4\,M$_\odot$, but drops precipitously beyond this and most
models predict that over 90\% of the cores remain neutron stars for
any EOS with a higher maximum rotating neutron star.  This strong
sensitivity to the EOS means that if we can observe these differences,
we have a strong probe of the neutron star equation of state.  We will
discuss the observational implications of these fits in
Section~\ref{sec:implications}.

Except at the boundary between most systems forming black holes and
most systems forming neutron stars, the differences between the
primary fates of our poplulation syntehsis models are fairly small
(few \%).  At the boundary, differences in the population of 10\% are
possible.  The primary uncertainty in our calculation is the lower limit 
on the neutron star formation mass.  For our lower NS formation limit of 1.1\,M$_\odot$, 
the Steiner2.4 (maximum non-rotating neutron star mass of 2.4\,M$_\odot$) 
marks the transition between mostly BH final states vs. mostly NS final states.  
If the mass is increased by 0.1\,M$_\odot$, the Steiner2.5 marks this transition.  
More importantly for our analysis is the transition EOS for systems that 
form BHs in less than 100\,ms.  For these systems, the transition is Steiner2.2, Steiner2.3 
respectively for our two mass limits.  The other population synthesis parameters do not 
produce much variation in these results.

\section{Implications}
\label{sec:implications}

\begin{figure}[!hbtp]
\centering
\includegraphics[width=\columnwidth]{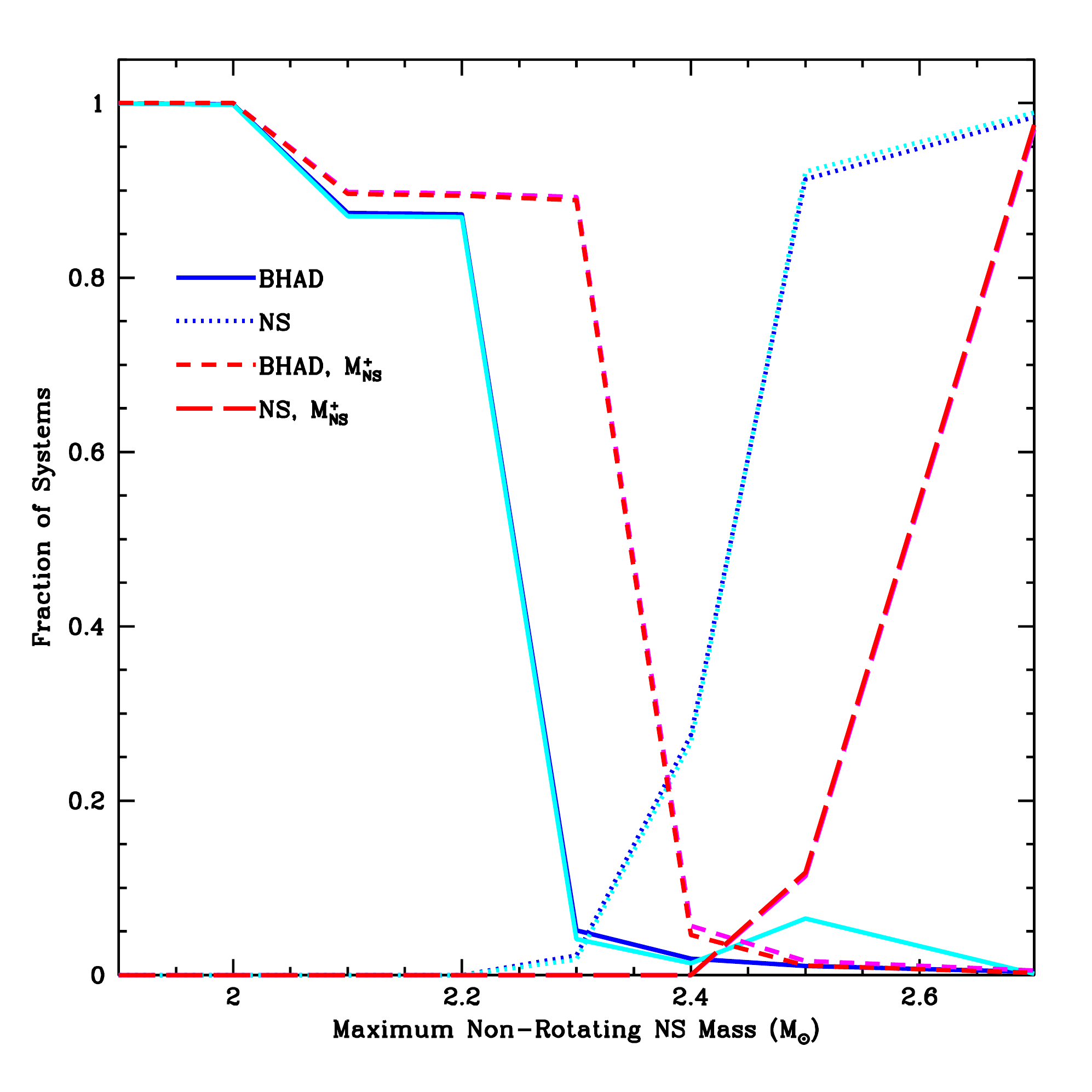}
\caption{Fraction of systems that produce standard BHAD GRBs and Neutron stars 
for our 4 basic population models:  standard NS masses (BHAD - solid, NS - dotted) 
for old (blue) and new (cyan) initial orbital parameters, NS masses increased 
by 0.1\,M$_\odot$ (BHAD - dashed, NS - long-dashed) for old (red) and new (magenta) initial 
orbital conditions.  The orbital conditions have virtually no effect on these results and 
the variations caused by other binary population synthesis parameters is also negligible.  The 
primary error in these estimates is that of the initial mass distribution of neutron stars, producing 
an equivalent error in the maximum non-rotating NS mass limit on the EOS.}
\label{fig:sum}
\end{figure}

In our analysis, we found that the fate of the merged system depends
sensitively on the neutron star equation of state.
Figure~\ref{fig:sum} shows the fraction of rapid ($t_{\rm acc}<100$\,ms)
and immediate collapse black holes as a function of the equation of
state.  If we can observationally distinguish between these fates, we
can use NS-NS mergers to place constraints on the equation of state.

For example, the current favored model for short-duration GRBs is the
merger of two neutron stars forming a black hole accretion disk.  If
the merged core forms a neutron star, neutrino-driven
winds~\citep{dessart09,perego14} drive outflows through which any jet
must penetrate to produce a gamma-ray burst.  Current studies argue
that this well-known baryon contamination problem can only be avoided
if the neutron star phase lasts less than $\sim
100$\,ms~\citep{murguiaberthier14}.  If this is true, GRBs are
produced in only a fraction of all merging systems.  If the maximum
mass is below 2.3-2.4 M$_\odot$, then the cores of most neutron star
mergers produce black holes within $\sim 100$\,ms of the merger.  For
these equations of state, most neutron star mergers are capable of
forming short-duration GRBs.

Above a maximum non-rotating neutron star mass of 2.3-2.4 M$_\odot$,
less than 4\% of all mergers produce rapid-collapse black hole systems
within 100\,ms.  If only such black hole systems form short-duration
GRBs, this would require a rate of mergers that is 50 times higher
than cannonical values.  Although possible, this strains our current
understanding of binary and stellar evolution, arguing strongly that
the maximum neutron star mass is below 2.3-2.4 M$_\odot$.  Advanced
LIGO will be able to place more stringent constraints on this merger
rate, and this measured merger rate will provide more firm constraints
on the maximum neutron star mass.

Unfortunately, constraining the NS-NS merger rate and its comparison
with short GRB rates may not directly constrain the equation of state.
BH-NS mergers can also produce short-duration GRBs and, in a small
subset of binary population synthesis scenarios, BH-NS merger rate is
on par with the NS-NS merger rate, e.g.\citep{dominik12}.  Although
it is possible to construct a scenario where BH-NS mergers dominate
the rate of NS-NS mergers, recall from Section 2.1 that it is likely
that most of BH-NS systems do not produce disks of sufficient size to
make GRBs~\citep{rosswog15a}.  If the connection between the short GRBs and BH-NS mergers
could be established, it will place strong constraints on binary
population synthesis models (in most scenarios, the NS-NS merger rate
is an order of magnitude higher than the BH-NS merger rate;
\cite{dominik12}.  It will also point to the EOS with maximum NS mass
over 2.3-2.4\,M$_\odot$ as such EOS allows to eliminate the formation
of short GRBs from NS-NS mergers via most accepted (BHAD) engine
model.

Alternatively, it is possible that the baryon loading from
neutrino-driven winds from neutron stars does not preclude the
possibility that cores that remain neutron stars can produce GRBs.
Late-time emission from short-duration GRBs either argues for some
extended power source (e.g. late-time accretion, magnetar) or jet
interaction\citep{holcomb14}.  The magnetar power source requires a long-lived neutron
star and has been shown, within the free parameters of the magnetar
model~\citep{rowlinson14}, to fit the late-time emission well.
Proponents of the magnetar model argue that there must be some way to
avoid the burying of the magnetic field and the baryon loading, e.g. a
high pressure, magnetic or radiation dominated, region along the axis
prevents baryon loading and allows the production of a relativistic
jet.

\begin{figure}[!hbtp]
\centering
\includegraphics[width=\columnwidth]{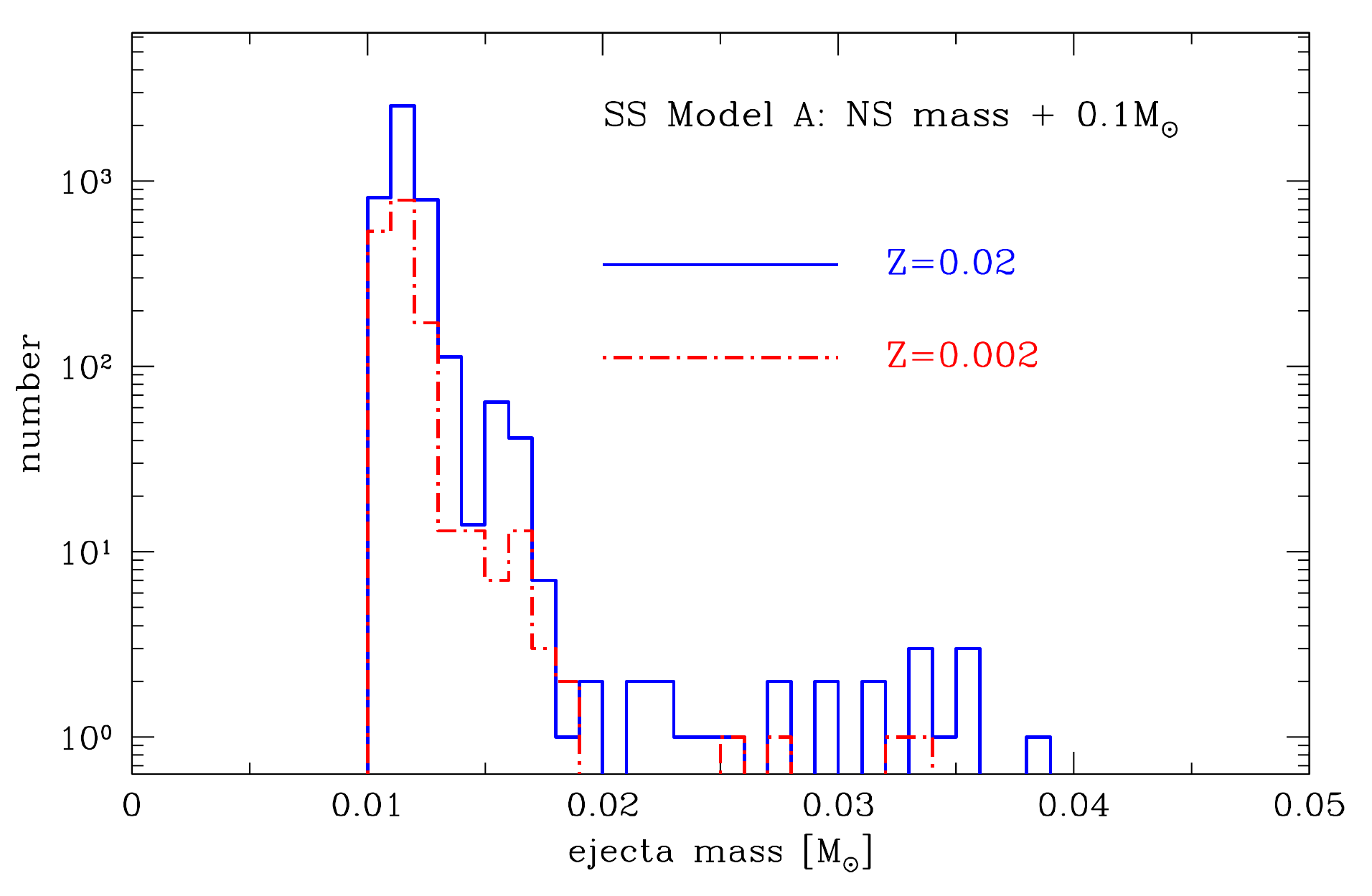}
\caption{The number of systems as function of ejecta mass in the tidal
  tails from these neutron star mergers for our increased NS models for 
both solar and 1/10th solar metallicity.  Note that althought the
  ejecta masses range from below 0.01\,M$_\odot$ to above
  0.03\,M$_\odot$, 75\% of the mergers produce ejecta with masses of
  $0.025\pm0.003$\,M$_\odot$.}
\label{fig:ejecta}
\end{figure}

To rule out these alternative solutions, we either need to refine our
theoretical models to prove the differences in the current
interpretations or we need additional observational diagnostics to
distinguish the above picture.  For example, it might be possible to
distinguish BH from NS formation using the gravitational wave
signal~\citep{chatziioannou15}, but this will likely require a very
strong signal from a nearby merger.  If a stable neutron star with
strong magnetic fields is formed, we would expect continued
pulsar-like activity, e.g. a soft gamma-ray repeater.  However, if we
only have LIGO localizations, it will be difficult to prove the
association of any activity with the merger, e.g.\citep{kelley10}.  The ejecta from these
models could allow us to distinguish these models.  The
optical/infra-red light produced by the decay of radioactive elements
can be used to distinguish between NS and BH fates.  The masses of the
dynamic ejecta that are shown in figure~\ref{fig:ejecta}) do not vary
dramatically in our models and, without alteration by the core, the
extreme neutron-richness of these ejecta leads to a composition that
almost exclusively consists of very massive nuclei with very large
opacities~\citep{kasen13} producing transients peaking in the
near-infrared (or
infrared)~\citep{roberts11,barnes13,tanaka13,grossman14,fontes15}.  However, the
transient observed post-merger is composed of emission produced by a
combination of this dynamic ejecta, neutrino driven winds from the
core (if the core does not collapse to a black hole) and ejecta from
the disk accretion.  Due to the longer exposure to the intense
neutrino background, the neutron fraction is reduced, altering the
composition of these ejecta and the revised opacities may allow
brighter transients in the near-infrared and optical\citep{perego14}.
Since this ejecta is much larger for NS systems than BH disk
systems\citep{fernandez15,just15}, we may be able to distinguish these
two systems by the wavelength of their electromagnetic emission.  If
so, we can use observations of neutron star mergers to place
constraints on dense matter equations of state.

{\bf Acknowledgments} This project was funded in part under the
auspices of the U.S. Dept. of Energy, and supported by its contract
W-7405-ENG-36 to Los Alamos National Laboratory. KB acknowledges
support from the Polish Science Foundation ``Master2013'' subsidy and
by the Polish NCN grant SANATA BIS.  This project orginated at a KITP
workshop supported by NSF Grant No. PHY11-25915. This work was
supported in part by the Simons Foundation and the hospitality of the
Aspen Center for Physics. The rotating neutron star simulations used
computational resources from the University of Tennessee and Oak Ridge
National Laboratory's Joint Institute for Computational Sciences.

{}

\begin{deluxetable}{llccccccc}
\tablewidth{0pt} 
\tablecaption{Merger Properties\label{tab:merger}\tablenotemark{a}}
\tablehead{\colhead{M$_{\rm NS 1}$} 
& \colhead{M$_{\rm NS 2} $}
& \colhead{M$_{\rm core} $\tablenotemark{b}}
& \colhead{S$_{\rm core} $\tablenotemark{b}}
& \colhead{J$_{\rm core} $\tablenotemark{b}}
& \colhead{M$_{\rm disk} $\tablenotemark{b}}
& \colhead{M$_{\rm ejected} $\tablenotemark{b}} 
& \colhead{$r_{\rm equator}$\tablenotemark{c}} 
& \colhead{$r_{\rm pole}$\tablenotemark{c}}\\
\colhead{$(M_\odot)$} 
& \colhead{$(M_\odot)$} 
& \colhead{$(M_\odot)$} 
& \colhead{$({\rm k_B/nuc})$}
& \colhead{$({\rm 10^{49} g cm^2 s^{-1}})$} 
& \colhead{$(M_\odot)$} 
& \colhead{$(M_\odot)$}
& \colhead{$({\rm km})$} 
& \colhead{$({\rm km})$} 
}

\startdata

1.0 & 1.0 & 1.51 & 0.91 & 1.43 & 0.48 & 0.0084 & 21.3 & 10.9 \\
1.2 & 1.0 & 1.58 & 0.93 & 1.12 & 0.59 & 0.029  & 20.9 & 12.9 \\
1.4 & 1.0 & 1.74 & 0.94 & 1.22 & 0.63 & 0.028  & 21.9 & 13.5 \\
1.6 & 1.0 & 1.86 & 1.07 & 1.20 & 0.71 & 0.030  & 18.2 & 13.6 \\
1.8 & 1.0 & 2.10 & 1.06 & 2.27 & 0.67 & 0.032  & 71.1 & 14.2\tablenotemark{d} \\
2.0 & 1.0 & 2.14 & 1.28 & 1.41 & 0.83 & 0.032  & 175.3 & 14.4\tablenotemark{d} \\
1.2 & 1.2 & 1.87 & 0.94 & 1.91 & 0.52 & 0.010  & 22.4 & 12.6 \\
1.4 & 1.2 & 2.01 & 0.95 & 1.92 & 0.57 & 0.022  & 24.0 & 13.4 \\
1.6 & 1.2 & 2.10 & 0.97 & 1.88 & 0.67 & 0.034  & 23.2 & 14.0 \\
1.8 & 1.2 & 2.19 & 1.19 & 1.87 & 0.77 & 0.036  & 21.9 & 14.9 \\
2.0 & 1.2 & 2.31 & 1.43 & 1.89 & 0.85 & 0.037  & 73.5 & 14.6\tablenotemark{d} \\
1.3 & 1.4 & 2.08 & 1.01 & 2.22 & 0.61 & 0.016  & 21.6 & 12.7 \\
1.3 & 1.4\tablenotemark{e} & 2.02 & 0.69 & 1.86 & 0.67 & 0.016  & 20.6 & 12.9 \\
1.4 & 1.4 & 2.29 & 0.96 & 3.11 & 0.50 & 0.012  & 25.2 & 11.6 \\
1.6 & 1.4 & 2.33 & 0.95 & 2.41 & 0.65 & 0.021  & 25.4 & 13.9 \\
1.8 & 1.4 & 2.40 & 1.00 & 2.15 & 0.76 & 0.038  & 25.2 & 14.3 \\
2.0 & 1.4 & 2.48 & 1.45 & 2.15 & 0.88 & 0.041  & 25.4 & 14.7 \\
1.6 & 1.6 & 2.55 & 1.03 & 3.07 & 0.64 & 0.013  & 23.7 & 13.6 \\
1.8 & 1.6 & 2.71 & 1.08 & 3.36 & 0.67 & 0.019  & 25.2 & 13.8 \\
2.0 & 1.6 & 2.70 & 1.36 & 2.76 & 0.86 & 0.042  & 24.5 & 14.6 \\
1.8 & 1.8 & 2.87 & 1.08 & 3.84 & 0.71 & 0.015  & 25.9 & 13.6 \\
2.0 & 1.8 & 3.04 & 1.28 & 4.11 & 0.74 & 0.019  & 26.4 & 14.2 \\
2.0 & 2.0 & 3.31 & 1.19 & 5.37 & 0.67 & 0.017  & 14.1 & 13.3 \\

\enddata

\tablenotetext{a}{All models from \citep{korobkin12} with the simulation physics from \citep{rosswog13}.}
\tablenotetext{b}{The core is defined by the post-merger material
  whose density is above $10^{14} {\rm g cm^{-3}}$ at the end of the
  calculation.}  
\tablenotetext{c}{$r_{\rm pole}$,$r_{\rm equator}$ correspond to the
  radius of the compact core along the rotation axis and equator
  respectively assuming the core reaches a uniformly-rotating,
  equilibrium state using the FSU2.1 equation of state.}
\tablenotetext{d}{For the most part, the angular momenta in these
  proto-neutron star is just below most limits for secular
  instabilities.  For these 3 systems, this is not the case.  It is
  possible that these instabilities can quickly shed angular momentum.}
\tablenotetext{e}{This model assumes co-rotation for the initial
  conditions.}  

\end{deluxetable}

\begin{deluxetable}{llccccccc}
\tablewidth{0pt}
\tablecaption{Remnant Fate\label{tab:nsfate}}
\tablehead{\colhead{M$_{\rm NS 1}$}
& \colhead{M$_{\rm NS 2}$}
& \colhead{Fate (eq.\ref{eq:nsm1})\tablenotemark{a}}
& \colhead{Fate (eq.\ref{eq:nsm2})\tablenotemark{a}}
& \colhead{Fate (eq.\ref{eq:nsm2})\tablenotemark{a}} 
& \colhead{Fate (eq.\ref{eq:nsm2})\tablenotemark{a}} 
& \colhead{Fate (eq.\ref{eq:nsm2})\tablenotemark{a}} 
& \colhead{Fate (eq.\ref{eq:nsm2})\tablenotemark{a}} 
& \colhead{Fate (eq.\ref{eq:nsm2})\tablenotemark{a}} \\
\colhead{$(M_\odot)$}
& \colhead{$(M_\odot)$} 
& \colhead{FSU2.1} 
& \colhead{$M_0=2.0$} 
& \colhead{$M_0=2.2$} 
& \colhead{$M_0=2.3$} 
& \colhead{$M_0=2.4$} 
& \colhead{$M_0=2.5$} 
& \colhead{$M_0=2.7$} }

\startdata
  1.0  &  1.0  & NS & NS & NS & NS & NS & NS & NS \\
  1.2  &  1.0  & BH$_{\rm spin}$  & BH$_{\rm acc}: {210 {\rm }}$ & NS & NS & NS & NS & NS \\
  1.4  &  1.0  & BH$_{\rm acc}: {180 {\rm }}$ & BH$_{\rm acc}: {40 {\rm }}$ & BH$_{\rm acc}: {220 {\rm }}$ & BH$_{\rm acc}: {1.2 {\rm s}}$ & NS  & NS & NS \\
  1.6  &  1.0  & BH$_{\rm acc}: {90{\rm }}$ & BH$_{\rm acc}: {90{\rm }}$ & BH$_{\rm acc}: {90 {\rm }}$ & BH$_{\rm acc}: {100 {\rm }}$ & BH$_{\rm acc}: {240 {\rm }}$ & BH$_{\rm acc}: {1.8 {\rm s}}$ & NS \\
  1.8  &  1.0  & BH$_{\rm acc}: {60{\rm }}$ & BH & BH$_{\rm acc}: {20{\rm }}$ & BH$_{\rm acc}: {90{\rm }}$ & BH$_{\rm acc}: {120{\rm }}$ & BH$_{\rm acc}: {360{\rm }}$ & BH$_{\rm spin}$ \\
  1.8  &  1.0\tablenotemark{b}  & BH & BH & BH & BH & 90 & 200 & BH$_{\rm spin}$ \\
  2.0  &  1.0  & BH$_{\rm acc}: {50{\rm }}$ & BH & BH & BH$_{\rm acc}: {50{\rm }}$ & BH$_{\rm acc}: {50{\rm }}$ & BH$_{\rm acc}: {50{\rm }}$ (BH) & BH$_{\rm acc}: {\rm 170 }$ \\
  2.0  &  1.0\tablenotemark{b}  & BH & BH & BH & BH & BH & BH & 90 \\
  1.2  &  1.2  & BH$_{\rm acc}: {500{\rm }}$ & BH$_{\rm acc}: {50{\rm }}$ & BH$_{\rm acc}: {450{\rm }}$ & BH$_{\rm acc}: >3{\rm s}$ & NS & NS & NS \\
  1.4  &  1.2  & BH$_{\rm acc}: {50{\rm }}$ & BH & BH$_{\rm acc}: {50{\rm }}$ & BH$_{\rm acc}: {280{\rm }}$ & BH$_{\rm acc}: {480{\rm }}$ & BH$_{\rm spin}$ & NS \\
  1.6  &  1.2  & BH$_{\rm acc}: {10{\rm }}$ & BH & BH$_{\rm acc}: {20{\rm }}$ & BH$_{\rm acc}: {30{\rm }}$ & BH$_{\rm acc}: {70{\rm }}$ & BH$_{\rm acc}: {200{\rm }}$ & BH$_{\rm spin}$ \\
  1.8  &  1.2  & BH$_{\rm acc}: {10{\rm }}$ & BH & BH$_{\rm acc}: {20{\rm }}$ & BH$_{\rm acc}: {30{\rm }}$ & BH$_{\rm acc}: {30{\rm }}$ & BH$_{\rm acc}: {30{\rm }}$ & BH$_{\rm acc}: {200{\rm }}$ \\
  2.0  &  1.2  & BH & BH & BH & BH & BH & BH$_{\rm acc}: {60{\rm }}$ & BH$_{\rm acc}: {\rm 60 }$ \\
  2.0  &  1.2\tablenotemark{b}  & BH & BH & BH & BH & BH & BH & BH$_{\rm acc}: {\rm 30}$ \\
  1.3  &  1.4  & BH$_{\rm acc}: {60{\rm }}$ & BH$_{\rm acc}: {60{\rm }}$ & BH$_{\rm acc}: {60{\rm }}$ & BH$_{\rm acc}: {70{\rm }}$ & BH$_{\rm acc}: {300{\rm }}$ & BH$_{\rm acc}: >3{\rm s}$ & NS \\
  1.3  &  1.4\tablenotemark{c}  & BH$_{\rm acc}: {\sim 3s}$ & BH$_{\rm acc}: {\sim 3s} $ & BH$_{\rm acc}: {\sim 3s} $ & BH$_{\rm acc}: {\sim 3s} $ & BH$_{\rm acc}: {\sim 3s} $ & BH$_{\rm acc}: {\sim 3s}$ & NS \\ 
  1.4  &  1.4  & BH$_{\rm acc}: {60{\rm }}$ & BH & BH$_{\rm acc}: {60{\rm }}$ & BH$_{\rm acc}: {60{\rm }}$ & BH$_{\rm acc}: {60{\rm }}$  & BH$_{\rm acc}: {490{\rm }}$ & BH$_{\rm spin}$ \\
  1.6  &  1.4  & BH & BH & BH & BH & BH$_{\rm acc}: {30{\rm }}$ & BH$_{\rm acc}: {60{\rm }}$ & BH$_{\rm acc}: {700{\rm }}$ \\ 
  1.8  &  1.4  & BH & BH & BH & BH & BH$_{\rm acc}: {20{\rm }}$ & BH$_{\rm acc}: {40{\rm }}$ & BH$_{\rm acc}: {80{\rm }}$ \\ 
  2.0  &  1.4  & BH & BH & BH & BH & BH & BH$_{\rm acc}: {40{\rm }}$ & BH$_{\rm acc}: {40{\rm }}$ \\  
  1.6  &  1.6  & BH & BH & BH & BH & BH & BH$_{\rm acc}: {30{\rm }}$ & BH$_{\rm acc}: {420{\rm }}$ \\ 
  1.8  &  1.6  & BH & BH & BH & BH & BH & BH$_{\rm acc}: {40{\rm }}$ & BH$_{\rm acc}: {130{\rm }}$ \\
  2.0  &  1.6  & BH & BH & BH & BH & BH & BH & BH$_{\rm acc}: {50{\rm }}$ \\ 
  1.8  &  1.8  & BH & BH & BH & BH & BH & BH & BH$_{\rm acc}: {70{\rm }}$ \\ 
  2.0  &  1.8  & BH & BH & BH & BH & BH & BH & BH$_{\rm acc}: {30{\rm }}$ \\ 
  2.0  &  2.0  & BH & BH & BH & BH & BH & BH & BH$_{\rm acc}: {40{\rm }}$ \\ 

\enddata

\tablenotetext{a}{The final fates are NS, the merged system forms a
  NSAD that, even after the accretion of a disk, remains a NS;
  BH$_{\rm acc}$, merged system initially forms a NSAD that, after
  some accretion, collapses to a BH forming a BHAD (the accretion time
  is given in ms unless otherwise specified); BH$_{\rm spin}$, merged
  system initially forms a NSAD that only collapses after the disk
  accretion and the loss of angular momentum; BH, the merged system
  collapses immediately to a BH, forming a BHAD; BH$_{\rm spin}$ (the
  merged system remains a NSAD until the disk accretion finishes after
  which spin down is able to decreases the maximum neutron star mass
  and allow it to collapse to a black hole.)}

\tablenotetext{b}{For these models, we consider the fate of the
  neutron star with a reduced rotation assuming the secular
  instabilities quickly reduce the angular momentum.  Any changes are
  noted in parantheses.  Note, however, that typical growth timescales
  for these systems are estimated to be above 100\,ms and it is more
  likely that the spin down occurs after the accretion phase.}

\tablenotetext{c}{This model assumes co-rotation for the initial
  conditions.}

\end{deluxetable}

\begin{deluxetable}{lcccccc}
\tablewidth{0pt} 
\tablecaption{Merger Fate Distributions\tablenotemark{a}\label{tab:popsynthfate}}
\tablehead{\colhead{Equation of State\tablenotemark{b}} 
& \colhead{BHAD GRB\tablenotemark{a}}
& \colhead{BH\tablenotemark{a}}
& \colhead{BH$_{\rm acc}$, $t_{\rm acc}<100$ms\tablenotemark{a}}
& \colhead{BH$_{\rm acc}$, $t_{\rm acc}>100$ms\tablenotemark{a}}
& \colhead{BH$_{\rm spin}$\tablenotemark{a}}
& \colhead{NS\tablenotemark{a}}
}

\startdata

FSU2.1\tablenotemark{c} & 0.8739 (0.8715)  & 0.0106 (0.0100)  & 0.8633 (0.8615)  & 0.1211 (0.1259)  & 0.0051 (0.0026)  & 0.0000 (0.0000) \\
Steiner2.0 & 0.9988 (1.0000)  & 0.8730 (0.8707)  & 0.1258 (0.1293)  & 0.0012 (0.0000)  & 0.0000 (0.0000)  & 0.0000 (0.0000) \\
Steiner2.2 & 0.8723 (0.8707)  & 0.0090 (0.0091)  & 0.8633 (0.8615)  & 0.1209 (0.1259)  & 0.0067 (0.0035)  & 0.0000 (0.0000) \\
Steiner2.3 & 0.0510 (0.0786)  & 0.0104 (0.0095)  & 0.0406 (0.0690)  & 0.8734 (0.9010)  & 0.0532 (0.0139)  & 0.0224 (0.0065) \\
Steiner2.4 & 0.0189 (0.0113)  & 0.0066 (0.0087)  & 0.0123 (0.0026)  & 0.1514 (0.0924)  & 0.5543 (0.4952)  & 0.2754 (0.4010) \\
Steiner2.5 & 0.0104 (0.0095)  & 0.0032 (0.0048)  & 0.0072 (0.0048)  & 0.0104 (0.0013)  & 0.0671 (0.0694)  & 0.9120 (0.9197) \\
Steiner2.7 & 0.0034 (0.0048)  & 0.0000 (0.0000)  & 0.0034 (0.0048)  & 0.0057 (0.0048)  & 0.0072 (0.0009)  & 0.9838 (0.9896) \\
& & & & & \\
FSU2.1\tablenotemark{d} & 0.8704 (0.8662)  & 0.0065 (0.0040)  & 0.8639 (0.8622)  & 0.1248 (0.1323)  & 0.0048 (0.0015)  & 0.0000 (0.0000) \\
Steiner2.0 & 0.9982 (1.0000)  & 0.8690 (0.8642)  & 0.1291 (0.1358)  & 0.0018 (0.0000)  & 0.0000 (0.0000)  & 0.0000 (0.0000) \\
Steiner2.2 & 0.8694 (0.8662)  & 0.0055 (0.0040)  & 0.8639 (0.8622)  & 0.1253 (0.1319)  & 0.0053 (0.0020)  & 0.0000 (0.0000) \\
Steiner2.3 & 0.0414 (0.0652)  & 0.0063 (0.0040)  & 0.0351 (0.0612)  & 0.8839 (0.9210)  & 0.0568 (0.0094)  & 0.0179 (0.0044) \\
Steiner2.4 & 0.0134 (0.0074)  & 0.0040 (0.0040)  & 0.0094 (0.0035)  & 0.1587 (0.0785)  & 0.5610 (0.5042)  & 0.2669 (0.4099) \\
Steiner2.5 & 0.0065 (0.0040)  & 0.0012 (0.0010)  & 0.0053 (0.0030)  & 0.0087 (0.0015)  & 0.0631 (0.0612)  & 0.9217 (0.9333) \\
Steiner2.7 & 0.0015 (0.0005)  & 0.0000 (0.0000)  & 0.0015 (0.0005)  & 0.0042 (0.0035)  & 0.0052 (0.0010)  & 0.9891 (0.9951) \\
& & & & & \\
FSU2.1\tablenotemark{e} & 0.8981 (0.8724)  & 0.0159 (0.0100)  & 0.8821 (0.8624)  & 0.1019 (0.1276)  & 0.0000 (0.0000)  & 0.0000 (0.0000) \\
Steiner2.0 & 1.0000 (1.0000)  & 0.8851 (0.8355)  & 0.1149 (0.1645)  & 0.0000 (0.0000)  & 0.0000 (0.0000)  & 0.0000 (0.0000) \\
Steiner2.2 & 0.8964 (0.8724)  & 0.0144 (0.0100)  & 0.8820 (0.8624)  & 0.1036 (0.1276)  & 0.0000 (0.0000)  & 0.0000 (0.0000) \\
Steiner2.3 & 0.8923 (0.8724)  & 0.0149 (0.0100)  & 0.8774 (0.8624)  & 0.1077 (0.1276)  & 0.0000 (0.0000)  & 0.0000 (0.0000) \\
Steiner2.4 & 0.0566 (0.0799)  & 0.0107 (0.0087)  & 0.0458 (0.0712)  & 0.8726 (0.8655)  & 0.0708 (0.0547)  & 0.0000 (0.0000) \\
Steiner2.5 & 0.0159 (0.0100)  & 0.0055 (0.0074)  & 0.0104 (0.0026)  & 0.0523 (0.0438)  & 0.8182 (0.7947)  & 0.1136 (0.1515) \\
Steiner2.7 & 0.0052 (0.0078)  & 0.0000 (0.0000)  & 0.0052 (0.0078)  & 0.0086 (0.0022)  & 0.0164 (0.0039)  & 0.9698 (0.9861) \\
& & & & & \\
FSU2.1\tablenotemark{f} & 0.8961 (0.8667)  & 0.0107 (0.0044)  & 0.8854 (0.8622)  & 0.1039 (0.1333)  & 0.0000 (0.0000)  & 0.0000 (0.0000) \\
Steiner2.0 & 1.0000 (1.0000)  & 0.8839 (0.8326)  & 0.1161 (0.1674)  & 0.0000 (0.0000)  & 0.0000 (0.0000)  & 0.0000 (0.0000) \\
Steiner2.2 & 0.8938 (0.8667)  & 0.0085 (0.0044)  & 0.8853 (0.8622)  & 0.1062 (0.1333)  & 0.0000 (0.0000)  & 0.0000 (0.0000) \\
Steiner2.3 & 0.8886 (0.8662)  & 0.0100 (0.0040)  & 0.8786 (0.8622)  & 0.1114 (0.1338)  & 0.0000 (0.0000)  & 0.0000 (0.0000) \\
Steiner2.4 & 0.0459 (0.0652)  & 0.0062 (0.0044)  & 0.0398 (0.0607)  & 0.8806 (0.8785)  & 0.0735 (0.0563)  & 0.0000 (0.0000) \\
Steiner2.5 & 0.0107 (0.0044)  & 0.0038 (0.0030)  & 0.0068 (0.0015)  & 0.0478 (0.0360)  & 0.8240 (0.7985)  & 0.1176 (0.1610) \\
Steiner2.7 & 0.0023 (0.0020)  & 0.0000 (0.0000)  & 0.0023 (0.0020)  & 0.0058 (0.0025)  & 0.0167 (0.0049)  & 0.9751 (0.9906) \\

\enddata

\tablenotetext{a}{We include 4 suites of simulations corresponding two
  different initial conditions: standard/old StarTrak initial
  conditions\citep{belczynski08b} and the new initial
  conditions~\citep{sana12} and two different initial compact remnant
  mass distributions:.  For each result, we include two values, one
  where all systems are included and one where it is assumed that the
  binary is destroyed if the system goes through a common envelope in
  the Hertzprung Gap.  For the remnant fates, we consider those
  systems that collapse immediately (within 3\,ms of the merger) to a
  BH (BH), systems that collapse to a BH within 100\,ms (BH$_{\rm
    acc}$, $t_{\rm acc}<100$ms), those that collapse to a BH through
  accretion taking more than 100\,ms (BH$_{\rm acc}$, $t_{\rm
    acc}>100$ms), those that collapse only after spin-down (BH$_{\rm
    spin}$), and those that remain neutron stars (NS).  The first
  column is the sum of the second and third columns (BH+BH$_{\rm
    acc}$, $t_{\rm acc}<100$ms) and represents those systems that can
  form the canonical black hole accretion disk GRB without significant
  baryonic contamination.}

\tablenotetext{b}{The equations of state correspond to the FSU2.1 equation of state with a maximum 
non-rotating neutron star mass of 2.1\,M$_\odot$ and a range of equations of state from our parameterized 
model (SteinerX) where ``X'' denotes the maximum non-rotating neutron star mass for that equation of 
state (in M$_\odot$).}
\tablenotetext{c}{This suite studies the old initial conditions with the standard NS mass distribution.}
\tablenotetext{d}{This suite studies the new conditions with the standard NS mass distribution.}
\tablenotetext{e}{This suite studies the old initial conditions with the enhanced ($+0.1$\,M$_\odot$) NS mass distribution.}
\tablenotetext{f}{This suite studies the new new conditions with the enhanced ($+0.1$\,M$_\odot$) NS mass distribution.}

\end{deluxetable}

\end{document}